\documentstyle [12pt] {article}

\begin{document}
\title {\large \bf  Representations\\ of The Quantum
Matrix Algebra $ M_{q,p}(2) $}

\author {Vahid Karimipour}
\date { }
\maketitle
\begin {center}
{\it  Department of Physics , Sharif University of Technology\\
P.O.Box 11365-9161 Tehran, Iran\\
Institute for studies in Theoretical Physics and Mathematics
\\ P.O.Box 19395-1795 Tehran, Iran \\
Email. Vahidka@irearn.bitnet}.
\end{center}
\vspace {10 mm}
\begin {abstract}
{It is shown that the finite dimensional irreducible
representaions of the quantum
matrix algebra $ M_{ q,p}(2) $ ( the coordinate ring of\ \
$ GL_{q,p}(2) $) exist
only when both q and p are roots of unity. In this case th
e space of states has
either the topology of a torus or a cylinder which may be thought of
as generalizations of cyclic representations.}
\end{abstract}
\noindent
\newpage
{\large \bf I. Introduction}\\

The representaion theory of quantized universal enveloping algebras [
 1-5 ] has been
extensively studied by many authors [ 6-10 ] and many beautiful
 results have been
obtained. Among these are the cyclic representaions which occur
when q is a root of unity.
However the representaion theory of the dual objects , that is the
 quantization of the
algebra of functions on the group ( Quantum Matrix Algebras ) has
not been systematically
studied . Only a few concrete representaions exist [ 11,12 ] . The
first attempt toward such
a goal has been reported in [ 13 ], where the irreducible finite
dimensional representations of $ M_q(2) $
were classified and it was shown that such irreducible representations
 exist only when q is a root of
unity . These representations were either cyclic or highest weight.

Although the representation theory of a multiparametric quantum
enveloping algebra is
trivial ( since the extra parameters appear only in the coproducts
 in the form of
a twisting [ 14 ] ) , the corresponding task for the multiparametric
quantum matrix group
is far from straightforward , since the extra parameters now appear in the
algebra itself.
As we will see this will lead to quite new features in the represen
representaions of the quantum matrix
algebra.

In this paper we study the representations of $ M_{q,p} ( 2) $ (
the coordinate ring of
$ GL_{q,p}(2) $) and classify its finite dimensional  irreducible
representations.
Our main reslults are the following:

i) finite dimensional irreducible representations exist only when
 both q and p are roots of unity.

ii) the space of states has the topology of a torus or that of a c
ylinder, depending on
the value of some parameters which define the representation.

iii) when none of the parameters q or p is a root of unity , the
states occupy  the sites of an
infinite rectangular lattice.

iv) when q is not a root of unity , but p is , the states occupy the sites
of an infinitely long cylinder.\\

{\large \bf II. the quantum matrix algebra $ GL_{q,p}(2) $}\\

This algebra [15] is generated by the entries of a matrix \ \
 $ T=\left( \begin{array}{ll} a&b \\ c&d \end{array} \right) $\\  and subject
to the relations:
\begin{equation} R T_1 T_2 = T_2 T_1 R \end{equation}
where R is a two parametric solution of Yang Baxter equation:

\begin{equation} R = \left( \begin{array}{llll} q &&& \\ & p && \\ & {
q-q^{-1}} & { p^{-1}}&\\ &&& {q} \end{array} \right) \end{equation}
The relations which follow from (1-2) are :
$$ ab = qp\ ba \hskip 2cm bd = { q\over p }\ db $$
\begin{equation} ac = { q\over p}\  ca \hskip 2cm cd = { qp }\  dc
\end {equation}
$$ bc = {1\over  p^2} \ cb \hskip 2cm ad - da =  p (q-q^{-1}) \ bc $$
Compared with $ GL_q(2) $ this quantum group has two particular
features which make its representation
theory quite different. The first is that the generators $ b $  and
$ c$  no longer commute and hence one
can not build the representation space from common eigenvectors of\ \ $ b $
  and $ c$  as in the work of [ 13 ].
The second is that the quantum determinant $ D = ad - qp\  bc $ is not
central, but satisfies the relations:
$$ D a = a D \hskip 2cm Db = p^2 bD $$
\begin{equation} D d = d D \hskip 2cm Dc = p^{-2} cD \end {equation}
The following relations can also be obtained by repeated
application of ( 3 )

$$ a^n d - d a^n = qp\ ( 1- q^{-2n}) \ a^{n-1} bc $$
\begin{equation} a d^n - d^ n a = { p\over q } ( q^{2n} - 1 )\
d^{n-1} bc \end {equation}
when $ q^n = p^n = 1 $ one sees from ( 3 ) and ( 5 ) that $ a^
n , b^n  , c^n  $ and  $ d^n $ are central.

In [13] both the commutativity of $ b $ and $ c $ and the
centrality of the quantum determinant have been
used to a large extent. In our case when the above facts are no longer
true , we must proceed in a different way.

Lets denote the product of $ b $ and $ c $ by M ( $ M = bc $ ).
 Then from (3) we find :
$$ M a = q^{-2} a M \hskip 2cm M b = p^2 b M $$
\begin{equation} M c = p^{-2} c M \hskip 2cm M d = q^2 d M\end
{equation}
Its clear from (4) that the operators M and D are
commuting:
\begin{equation} M D = D M  \end {equation}
We will use these commuting operators in the following sections to
build up the irreps. of $ M_{q,p}(2) $.

\vskip 1cm
{\large \bf III. Finite Dimensional Irreducible Representations }\\

Following [13] we call those $ M_{q,p}(2)$ modules in which one or
more of the generators identically vanish,
trivial modules . In these cases the representation reduces to that
of a simpler algebra. Clearly the interesting
representations are nontrivial ones to which we restrict ourselves in
 the rest of this paper.
The following lemma [13] establishes the condition for nontriviality
of the representation.\\

{\bf Lemma 1}: Let V be a vector space . Then an irreducible
representation
$ \rho : M_{q,p}(2)\longrightarrow End (V) $  is trivial if $ b $ or $ c
$ have  zero
eigenvalue in their spectrum.\\

{\bf Proof} : We follow a slightly different line of reasoning which is
simpler
 than that of [13]. Without loss of generality , lets assume that \ $ K_b
 \equiv Ker \ b \ne\  \{ 0\}\  $ , then since  $ K_b $  is a subspace
of V , we can choose
a basis for it like :$ \{ e_1 , . . .  e_ m \} $ . From eq. (3) we see that \
$ a K_ b \ , \ c K_b\  , $  and \ $  d K_b\  $ are all subspaces of $ K_b $ .
 Therefore the vectors $ \{ e_i\} $  transform among themselves
under the action of
$ a , b , c $ and  $ d $  and hence  $ K_b $  is an invariant subspace
 of V . Since the representation is irreducible $ K_b = V $ . Therefore
$ b  K_b = b  V = 0 $  and the representation is trivial.
Hereafter we assume that  $ K_b = K_c = \{ 0 \} $ .\\

{\bf Lemma 2}:  A finite dimensional irreducible $M_{q,p}(2) $  module
exists only when
both $ q $ and $ p $  are roots of unity.\\

{\bf Proof}: Let $ v_0$  be a common eigenvector of $ M $ and $ D $ .
\begin{equation} M v_0  = \mu v_ 0 \hskip 2cm D v_0  = \lambda v_0
\end {equation}
Then from ( 3,4, and 6 ) one sees that the string of states $ v_n
 \equiv d^n v_0  $ satisfy:
\begin{equation} M v_n = q^{2n} \mu   v_n \hskip 2cm D v_n = \lambda
v_n \end{equation}
For the parameter q  we adopt the reasoning of ref.[13].
To have finite dimensional representations one must have $ d^l v_0 = 0  $
 for some $ l $
while all the vectors $ v_n $ for $ n<l $ are independent.Consider the
 string of states  $ a^m u_0 $ where $ u_0 = v_{l-1} $. Again one must have
 $ a^ {l'} u_0 = 0 $ but $ a^{ l'-1} u_0 \ne 0 $. Then one will have
$$ 0 = da^{l'}u_0 = {qp}( q^{-2l'}-1)\mu a^{l'-1}u_0 $$
which means that q must be a root of unity.
On the other hand, suppose that  $ p $ is not a root of unity. then
the string of states
$ u_n = b^ n v_0 $  satisfy:
$$ M u_n = p^{2n}\mu\  u_n \hskip 2cm D u_n = p^{2n}\lambda u_n $$
If $ p $ not a root of unity  all the above states will be independent and
 the representation can not be finite.

Hereafter we set $ q^r = p^r = 1 $  ( note: $ q $ and $  p $ may be
different roots of unity ,  i.e:  \ $  q^ {r_1} = p^{r_2 }  = 1 $ . We set\
  $ r $\ to be the least common
multiplier of $ r_1 $  and $ r_2 $  ).

In this case $ a^r  , b^r , c^r  $ and $ d^ r $  are central and on V
 we set them equal to $ \eta_a , \eta_b ,
\eta _c $  and  $ \eta _d $ respectively. Clearly   $ \eta _b $ and  $
 \eta_c $  are both different from zero, otherwise $ K_b $ and $ K_c $
  will have nonzero elements.

We now classify the finite dimesnsional irreducible representations
of $ M_{q,p}(2) $ which fall into three classes
depending on the values of $ \eta_a $  and $ \eta_d $ .
\vskip 1cm
{ \bf A. Toroidal Representations ( $ \eta_a \ne 0 \ne \eta_d $ )}\\

We denote the vector $ v_0 $ introduced in (8) by $ \vert 0 , 0 > $  and c
onsider the lattice of states $W$ (fig. 1):

\begin{equation} W = \{ \vert l , n > \equiv b^l d^n \vert 0 , 0 > \hskip 1cm
0\leq l , n \leq r-1 \} \end {equation}
These states are the common eigenvectors of $ M $ and $ D $ ( see 3-6).
\begin{equation} M \vert l ,n > = p^{2l}q^{2n} \mu\vert l,n> \hskip
1cm D\vert l ,n > = p^{2l}\lambda \vert l , n > \end {equation}
We define the action of $ a $  and $ c $  on the state $ \vert 0 , 0
> $  as follows:
\begin{equation} a \vert 0 , 0 > = \alpha_0 \vert 0 , r-1 > \hskip 1
cm c\vert 0, 0 > = \gamma_0  \vert r-1 , 0 >\end {equation}

Then we have :

{\bf Theorem 3 } : The following defines an irreducible representation of
$ M_{q,p}(2) $.\\
i) $ b\vert l , n> = \vert l+1 , n> \hskip 2cm b\vert r-1 , n> =
\eta _b\vert 0 , n > $\\

ii)$ d\vert l , n> = ({p\over q })^l\vert l , n+1 > \hskip 2cm  d\vert
 l , r-1> = ( { p\over q})^l \eta_d \vert l , 0 >  $\\

iii)$ c\vert l , n> = p^{2l}q^{2n}\eta_b \gamma_0 \vert l-1 , n >\hskip
 1cm  c\vert 0 , n > = q^{2n} \gamma_0 \vert r-1 , n >  $\\
iv)$ a\vert l , n> = (qp)^l ( \alpha_0 \eta_ d + {p\over q } ( q^{2n}-
1)\gamma_0 \eta_b) \vert l , n-1 >$                        \\

$ a\vert l , 0 > = ( qp)^l \alpha_0 \vert l , r-1 > $

{\bf Proof}:  i)  and  ii) are obvious . we give an explicit verification
of iii). iv )is obtained by
straightforward manipulations.
Acting with $ c $ on the state $ \vert l , n > $  and using the commutation
 relataions ( 3 ) we find :
$$ c \vert l,n > = c b^l d^n \vert 0, 0 > = p^{2l} ( qp )^n b^l d^
n c \vert 0 , 0 >  $$
using ( 12 ) and the definition of states ( 10 ) we arrive at :
$$ c \vert l,n > = p^{2l} ( qp )^n b^l d^n \gamma_0  \vert r-1 , 0 >
= p^{2l} ( qp )^n \gamma_0 b^l d^n b^{r-1} \vert 0 , 0 >
$$ againg using the commutation relations and the fact that   $ b^r =
 \eta_b $
we finally find:    $$ c \vert l ,n > = p^{2l} ( qp )^n \gamma_0 (
{p\over q})^{n(r-1)}b^{l-1} d^n  \vert 0,0 > = p^{2l} q^{2n} \eta_b
\gamma_0 \vert l-1 , n > $$
The second part of iii) is proved similarly :

$$ c\vert 0,n > = c  d^n \vert 0, 0 > =  ( qp )^n  d^n c \vert 0 ,0 >  $$
$$= ( qp )^n d^n \gamma_0  \vert r-1 , 0 >
= ( qp )^n \gamma_0  d^n b^{r-1} \vert 0 , 0 >
$$ $$ = ( qp )^n \gamma_0 ( {p\over q})^{n(r-1)}b^{r-1} d^n \vert 0,0
 > = q^{2n} \gamma_0 \vert r-1 , n > $$
It is interesting to note that the space of states has the topology of
 a torus
 ( see fig. 2 where the actions of all the generators
are shown graphically).

The dimension of this representation is $ r^2 $. To prove that it is
 the  only irreducible representation in this case , we
note that the dimension of V can not be greater than $ r^ 2 $  since
 otherwise the above lattice of states which is based on a single
common eigenvector of $ M $  and $ D $ will provide an invariant subspace
 which contradicts the irreducibility  of the representation.
The dimension of V can not be less that $ r^ 2 $ either  since then one of
 the strings of states $ d^n \vert 0 , 0 > $  or $ b^n \vert 0 , 0 > $
  must terminate for
some value of $ n $  less than $ r $ .( i.e:$ d^n \vert 0 , 0 > = 0 , n <
 r $) This then means that
$$ \eta_d\vert 0,0 > = d^ {r-n} ( d^n \vert 0, 0 >) = 0 $$
which contradicts the assumption of $ \eta_d \ne 0 $.\\

{\bf Remark }: The parameters $ \alpha_0 \ \  \gamma_0 \ \ \lambda $  \
an d $ \mu $ are not
independent of $ \eta_a ,\eta_b ,\eta_c  $ and $ \eta_d $. The following
 relations exist among them:
\begin{equation} \eta_c = \gamma_0^r \eta_b^{r-1} \hskip 3cm \mu =  \gamma
_0 \eta_b  \end {equation}
\begin{equation} \lambda = \alpha_0 \eta_d - {p\over q} \mu \hskip 2 cm
\eta_a = \alpha_0 \prod_{i=1}^{r-1} ( \lambda + { p\over q } \mu  q^{2i})
\end {equation}
The proof of these formulas is given in the appendix. We now turn to the
second kind of representations.
\\

{\bf B. Cylindrical Representations $ (\eta_a = \eta_d = 0) $}\\

In this case the representation in theorem 3 is modified as follows:

\begin{equation} d \vert l , r-1 > = 0 \ \ \ \ \ \ \ \ a \vert l , 0 >
= 0 \end{equation}
\begin{equation} a \vert l,n > = ( qp)^l( {p\over q} ( q^{2n} - 1 )
\gamma _0 \eta _b ) \vert l, n-1> \end{equation}

all the other relations remain intact. This kind of representation may be
called a cylindrical representation and may be thought of as
a truncated form of the toroidal representation.( fig. 3 )

C) Where only one of the parameters ( say $ \eta_a $ ) is zero , the
relations ( 15-16 ) are modified as follows:
\begin{equation} d \vert l , r-1 > = \eta_a \vert l ,0 > \ \ \ \ \ \
\ \ a \vert l , 0 > = 0 \end {equation}
\begin{equation} a \vert l,n > = ( qp)^l( {p\over q} ( q^{2n} - 1 )
 \gamma_0 \eta_b ) \vert l, n-1> \end{equation}
This then may be thought of as a semitoroidal representaion in which
$ d $ traverses completely one of the cycles of the torus while $ a $ dose
 not.
\vskip 1cm

{\large \bf IV. Infinite Dimensional Representations }\\

In theorem 3 one can relax the conditions on the right hand side
. Then one can check easily that the left hand side equations
define an infinite
dimensional representation of $ M_{q,p}(2) $ on the two dimensional
lattice :
$ W = \{ \vert l,n > \ \ \ \ \ -\infty < l , n < \infty \} $
$$ b\vert l , n> = \vert l+1 , n >$$
$$ d\vert l , n> = ({p\over q })^l\vert l , n+1 > $$
\begin{equation} c\vert l , n> = p^{2l}q^{2n}\mu \vert l-1 , n >
\end{equation}
$$ a\vert l , n> = (qp)^l ( \lambda + ({p\over q }\mu)  q^{2n}) \vert
l ,n-1> $$

{\bf Note}: The states of this representation are not necessarily built up
on a vaccum  ( i.e $ \vert l ,n > \ne b^l d^n \vert 0 , 0 > $ ).

If both $ q $  and $ p $  are roots of unity ( $ q^r = p^r = 1 $ ) then
one can consistently identify the states as follows:

$$ \vert l,n> \equiv \vert l+r , n > \ \ \ \ \ \ \ \ \vert l,n>
\equiv \vert l,n+r > $$
Representation ( 19 ) will induce a Toroidal representation on the
equivalence class of these states.

If only $ p $ is a root of unity , then one  can consistently identify
 the states as follows :

$$ \vert l+r , n > \equiv q^{nr} \vert l,n > $$
Eqs. (19) then induces  a cylindrical representation on the the
equivalence of these states.
By setting $ \lambda = -{ p\over q } \mu  $ one can also obtain a
lowest weight module.
\vskip 1cm

{\large \bf Discussion }\\

In the classical limit (q = p = 1),the coordinate ring of $ GL_{q,p}(2
) $ degenerates
into a free abelian algebra whose irreducible finite dimensional
representations
are one dimensional where the generators $ a, b, c, $ and $ d $ are
represented by
by the pure numbers $ \eta_a , \eta_b , \eta_c , $ and $ \eta_d  $
respectively.
This limit is obtained from the representations in this paper by noting
that when
$ r = 1 $ , the lattice of states in fig. 1 has only one single state,
namely $ \vert 0,0 > $
and from theorem I we have:
$$ a\vert 0, 0 > = \alpha_0 \vert 0, 0 >\ \ \ \ \ \ \ \ \  b\vert
0, 0 > = \eta _b \vert 0, 0 >$$
$$ c\vert 0, 0 > = \gamma_0 \vert 0, 0 >\hskip 2cm    d\vert 0, 0 > =
\eta_d \vert 0, 0 > $$
{}From eqs.(14-15) we also obtain $ \gamma_0 = \eta_c $ and $ \alpha_0 =
\eta_a $
which proves the assertion.

What appears to be very interesting about the representation theory of
 quantum
matrix algebras compared with those of the quantized universal
enveloping algebras
is that in the latter case the classical limit is a lie algebra and one
 can use
the decomposition of the root space of the lie algebra into the Cartan
subalgebra
and positive and negative root system for building up the represetation.
 Recall that
this decomposition remains essentially intact in the process of quantization
{}.
This then leads to a paralellism between the represetation theory in
the deformed
and the undeformed case.
However for the case of quantum matrix algebras such a decomposition and
the resulting paralellism does not
exist.One expects that completely new feature
s arise
in their representation theory.(see for example [16] and [17]). \\
\vspace {1cm}             \\
 {\large \bf Appendix:  Proof of eqs. (13) and (14) }

Repeated use of theorem 3 gives the following:
$$ c^{r-1} \vert r-1,0> =(\eta_b \gamma_0)^{r-1} \vert 0 , 0 > $$
Acting by $ c $ on both sides we obtain
$$ \eta_c \vert r-1 , 0 > = (\eta_b \gamma_0)^{r-1}\gamma_0 \vert r-
1 , 0 > $$
where we have used (12) . Comparison of both sides gives the
first relation of (13).
To obtain the second relation of ( 13 ) we note that $$ \mu \vert 0, 0
 >= M \vert 0 , 0> = bc \vert 0 , 0 > =
b ( \gamma_0 \vert r-1 , 0 > ) = \gamma_0 \eta_b \vert 0 , 0 > $$
To prove the first relation of (14) we act on the state $ \vert 0,0 > $
with  $ D = ad - qp M $  and use theorem 3 .
Finally we note that:
$$ a^{r-1} \vert 0 , r-1 > =  \prod_{i=1}^{r-1} ( \lambda +({p\over q }
\mu) q^{2i})\vert 0 , 0 > $$

Acting on both sides with  $ a $ and using (12) gives (14).
\vskip 1cm

\newpage
{\large \bf References }
\begin{enumerate}
\item {1-} V. G. Drinfeld , Proceeding of the ICM ( Berekely,
Berkeley, CA, 1986) p.798.
\item {2-} M. Jimbo , Lett. Math. Phys. 10,63 (1985) ; 11, 247 (1986)
\item {3-} N. Yu. Reshetikhin , L. Takhtajan , L. D. Faddeev  : Alg.
Anal. 1 (1989) 178-206 (in russian)
\item {4-} S. Woronowics, Commun. Math. Phys. 111,613 (1987); 130,387 (1
990)
\item {5-} Y. Manin , CRM preprint (1988)
\item {6-} G. Luszig , Adv. Math. 70 , 237 (1988) ;Contemp. Math. 82
 , 59 (1989)
\item {7-}Rosso M.: Commun. Math. Phys. 117,581 (1988) 124, 307 (1989)
\item {8-}C. P. Sun J. F. Lu and  M. L. Ge :  J.Phys A. 23 ,L 1190 (1990)
\item {9-} N. Yu. Reshetikhin , LOMI preprints , E-4 and E-11 (1987)
\item {10-} R. P. Roche and D. Arnaudon , Lett. Math. Phys. 17, 295 (1989)
\item {11-} E. G. Floratos , Preprint CERn - TH , 5482/89
\item {12-} J. Weyers, Preprint , CERN-TH 5633/90
\item {13-}M. L. Ge and X. F. Liu , and C. P. Sun : J. Math. Phys 33
(7)2541 (1992)
\item {14-} Yu. N. Reshetikhin ;  Lett. Math. Phys. ( 1990 )
\item {15-} A. Sudbery , J.
Phys. A., 23 (1990) L697
; A. Schirrmacher,  Z. Phys. C 50, 321,(1991) , A. Schirrmacher , J. Wess
, B. Zumino, Z. Phys. C 49 ,317, (1991).
\item {16-} V. Karimipour; Representations of the Coordinate Ring of $ GL
_q(3) $,
IPM preprint 93-010 , Tehran .To appear in Lett. Math. Phys.
\item {17-} V. Karimipour; Representations of the Coordinate Ring of $
GL_q(n) $,
in preparation
\end{enumerate}
\vfil\break

FIGURE CAPTION

Fig. 3 : The Lattice of States W

Fig. 2 : Toroidal Representation

Fig. 3 : Cylindrical Representation
\end{document}